\pdfoutput=1
\documentclass[twocolumn,aps,prb]{revtex4-2}
\usepackage{graphicx}
\usepackage{amsmath}
\usepackage{float}
\usepackage{subcaption}
\usepackage{tabularx}
\begin{document}

\title{Ab Initio strain-dependent vibrational analysis of different silicene structures for 2D devices}

\author{Archishman Gupta}%
 \email{Contact author: archishman.gupta.cd.ece21@itbhu.ac.in}
 \author{Ankit Arora}%
 \email{Contact author: ankit.ece@iitbhu.ac.in}
\affiliation{%
 Indian Institute of Technology (BHU) Varanasi, India.\\}%

\date{\today}

\begin{abstract}
We report first-principles phonon dynamics studies in isotropically strained mono-layer silicene films of various topologies by density functional theory and density functional perturbation theory methods. The strain introduced in the sheets are of both tensile and compressive nature and its effect on the thermal transport properties are analysed using the Boltzmann Transport Equation (BTE) reported. We performed electronic and vibrational analysis of each silicene topology and concluded that hhk-silicene is a potential candidate for next generation electro-thermal devices due its molecular stability, bulk-Si like bandgap and tolerable thermal performance that can be engineered by strain. The novel hhk-silicene topology offers excellent thermal and electronic capabilities that are sensitive to strain along with the band gap following a near-linear monotonically decreasing trend making it suitable for strain-tronics applications.

\end{abstract}

\maketitle

\section{Introduction}
The advent of graphene has catalyzed a paradigm shift in materials science, unveiling the extraordinary potential of two-dimensional (2D) systems for next-generation technologies. Among these, silicene — a silicon-based analogue of graphene—has emerged as a compelling candidate due to its compatibility with established silicon electronics and unique electronic properties. Unlike graphene’s flat lattice, silicene adopts a buckled geometry owing to its sp-3 hybridization compared to graphene's sp-2 hybridization.

Despite this promise, practical deployment of silicene in nano-electronics has been impeded by inherent challenges, including structural instability and the absence of a non-zero bandgap in its pristine form. Recent explorations into diverse silicene topologies, ranging from planar to buckled to \textit{kagome} configurations, aim to address these limitations. However, reconciling stability, electronic functionality, and thermal performance remains a critical hurdle.

Strain engineering has proven transformative in tailoring the properties of 2D materials. By applying isotropic or anisotropic strain, electronic band structures and phonon dynamics can be precisely modulated, offering avenues to optimize materials for flexible electronics, thermoelectrics, and strain-tronic devices. Prior studies on silicene have predominantly focused on strain-induced phononic observations, such as phonon mode contributions, yet the effects on thermal transport and vibrational properties—particularly across distinct topological structures of silicene remain poorly understood. This knowledge gap is pivotal, as thermal management is integral to the reliability and efficiency of nano-electronic systems.

Here, we present a comprehensive first-principles investigation of isotropically strained monolayer silicene across multiple topologies, employing density functional theory (DFT) and density functional perturbation theory (DFPT) to elucidate electronic and vibrational responses. Thermal transport properties are further analyzed via the Boltzmann Transport Equation (BTE). Central to this study is the examination of a novel silicene topology, termed hhk-silicene in Sang et al \cite{sangSemiconductingSiliceneTwoDimensional2021}, which exhibits exceptional molecular stability and a bulk silicon-like bandgap. Our results demonstrate that isotropic strain—both tensile and compressive—induces a near-linear reduction in the electronic bandgap while significantly altering phonon dispersion and thermal conductivity. These dual electronic and thermal responses position hhk-silicene as a versatile platform for electro-thermal devices, where strain-dependent thermal switching and bandgap engineering can be synergistically exploited.

By bridging the gap between electronic and thermal analyses under strain, this work helps advance the fundamental understanding of 2D silicon-based systems.

\section{Simulation Methods}

First-principles calculations were conducted using density functional theory (DFT) as implemented in the Quantum Espresso (QE) package. The projector-augmented-wave (PAW) method and the Perdew-Burke–Ernzerhof (PBE) exchange-correlation functional were employed to model electron interactions. To simulate monolayer silicenes, a vacuum layer of 20 \r{A} was introduced along the out-of-plane (z-axis) direction to mitigate periodic boundary interactions. A plane-wave basis set with a cutoff energy of 30 Ry and a Monkhorst-Pack k-point grid of $11 \times 11 \times 1$ were used to ensure convergence. 
Geometric optimizations were performed until atomic forces and electronic energies converged below thresholds of $10^{-7}$ eV/\r{A}.

Phonon transport properties were evaluated by solving the phonon Boltzmann transport equation (pBTE) using the Phono3py package. This requires harmonic and anharmonic interatomic force constants (IFCs), which were derived from DFT calculations. Harmonic and anharmonic IFCs were obtained using the Phonopy code with a $2 \times 2 \times 1$ supercell. The pBTE was iteratively solved in Phono3py for calculating the lattice thermal conductivities of the silicenes. Note that four-phonon scattering processes were excluded in this study, as their influence is typically significant only at elevated temperatures or in materials exhibiting exceptionally high lattice thermal conductivity.

\section{Results}

The structural, electronic, and thermal properties of silicene were analyzed across Flat, Buckled, Kagome, and hhk topologies using DFT and DFPT; with hhk-silicene emerging as the most promising due to its strain-sensitive bandgap and tunable thermal conductivity for strain-tronics and electro-thermal applications. The crystal structure of hhk-silicene, shown in FIG. \ref{fig:1}(d), is formed by substituting one site in the honeycomb lattice with a kagome triangle, as proposed by Sang et al. \cite{sangSemiconductingSiliceneTwoDimensional2021}, leading to electronic bands that fall between Dirac-cone and flat dispersion, opening a bandgap near the Fermi level. The relaxed lattice constant is 5.86 \r{A}, closely matching previous reports, and the primitive cell contains four silicon atoms, where three kagome-$Si_3$ triangles connect via a bridge Si atom, resulting in sp² and sp³ hybridization. The Brillouin zone path follows $\Gamma-M-K-\Gamma$, and estimate an indirect bandgap of 1.18 eV, which, compared to larger bandgaps in TMDs like $MoS_2$ (1.86 eV), suggests better electronic transport properties. Additionally, the calculated phonon spectrum, shown in FIG. \ref{fig:1}(l), confirms the dynamical stability of hhk-silicene, as it lacks negative frequency modes.

\subsection{Electronic Properties}

\begin{figure*}[t!]

    \includegraphics[width=15cm]{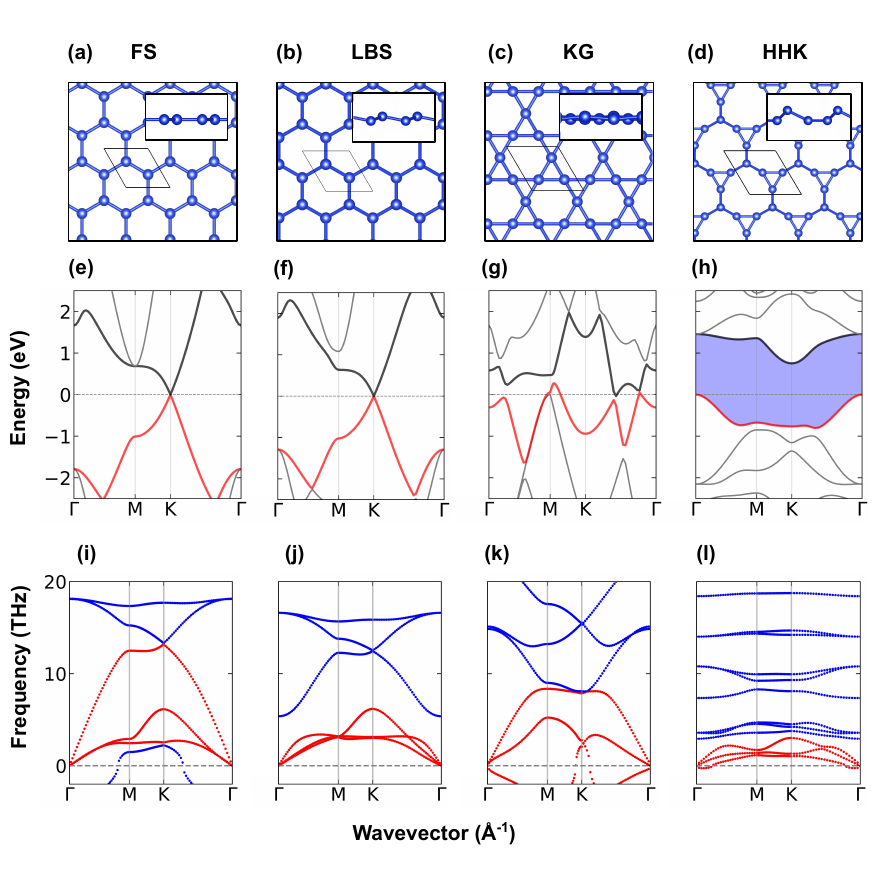}
    \captionsetup{justification=raggedright,singlelinecheck=false}
    \caption{
    (a)-(d) : represents the 2D lattice structures of Flat, Low buckled, Kagome and Hybrid honeycomb Kagome topologies. (e) - (h) : Electronic structures, with only HHK showing non zero bandgap. (i) - (l) : Phonon dispersions of FS, LBS, KG and HHK silicenes.}
    \label{fig:1}
 
\end{figure*}

The electronic properties of silicene were analyzed for four different topologies: Flat, Buckled, Kagome, and Hybrid Honeycomb-Kagome (hhk) structures. The computed band structures (Fig. \ref{fig:1}(e-h)) reveal significant differences among these configurations. Flat and buckled silicenes exhibits a zero-bandgap semimetallic behavior, making it unsuitable for semiconductor applications due to the lack of a controllable bandgap. Kagome silicene moves away from the dirac cone topology however it still lacks a finite bandgap needed for semiconducting applications. Most notably, hhk-silicene displays a bulk-Si-like bandgap, suggesting better compatibility with conventional silicon electronics.

\begin{figure*}[t!]

        \begin{center}
            
            \includegraphics[width=15cm]{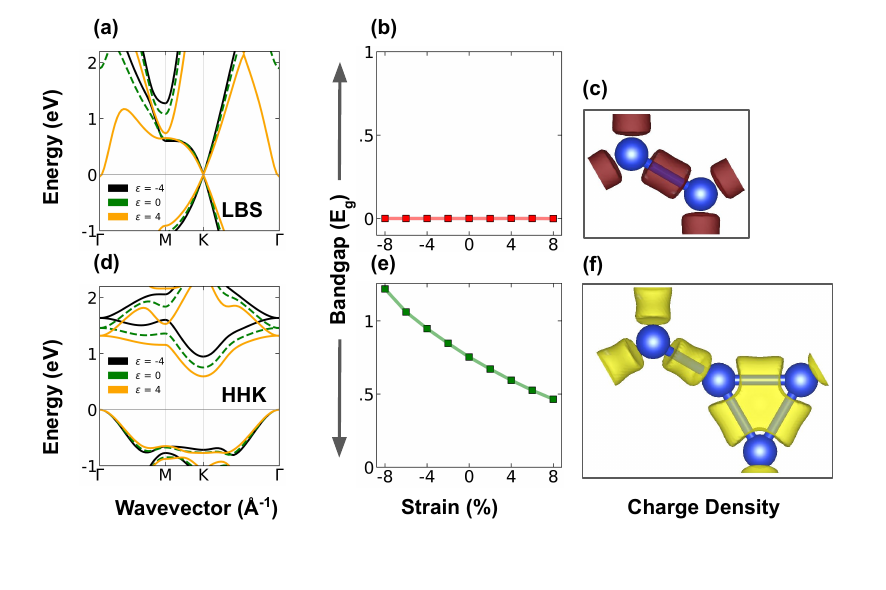}
            \caption{ (a) : The valence band maxima (VBM) and conduction band minima (CBM) of LBS are plotted for 3 values of strain. (b): LBS bandgap as a function of strain and (c) shows the localised charge density for a single lbs-unit cell (d) : The valence band maxima (VBM) and conduction band minima (CBM) of HHK are plotted for 3 values of strain. (e): HHK bandgap as a function of strain and (f) shows the localised charge density for a single hhk-unit cell}
            \label{fig:2}
        \end{center}
\end{figure*}

After performing relax calculations, the cell constants in their most stable configurations were found to be 3.85, 3.86, 4.45 and 5.86 \r{A} for FS, LBS, KG and HHK respectively.
Further examination of strain-dependent bandgap evolution in hhk-silicene (Fig. \ref{fig:2}(e) reveals that under strain, the hhk bandgap experiences a monotonic decrease with strain following a linear trend (going from 1.1 to 0.5 eV when put under strain of -8 to +8 \%). This linear bandgap modulation under strain makes hhk-silicene a promising material for strain-tronics, where mechanical strain can dynamically control electronic properties. Such behavior allows for its application in adaptive and energy-efficient electronic devices.

The charge density distributions (Fig. \ref{fig:2}(f)) provide insight into the electron localization and bonding characteristics of these hhk-silicene, hhk-silicene maintains a well-distributed and localised charge density, contributing to its structural stability and superior electronic performance.

Its strain-tunable bandgap makes it particularly suitable for strain-engineered electronics, flexible transistors, and thermal management systems. Additionally, its bulk-Si-like bandgap enhances its compatibility with existing silicon-based semiconductor technology, offering a seamless integration pathway. The observed strain-dependent bandgap tuning opens up new possibilities for energy-efficient computing and adaptive electronic systems, further solidifying hhk-silicene as a prime candidate for flexible, high-performance nanoelectronics.

\subsection{Vibrational Properties}
The phonon dispersion curves for different silicene topologies - Flat, Low-Buckled (LBS), Kagome, and hhk-Silicene are presented in the following figures. Fundamental vibrational properties provide insights into the structural stability, transport and lattice dynamics of each structure.

\subsubsection{Phonon dispersion}

For Flat Silicene (Fig. \ref{fig:1}(i)), the phonon spectrum exhibits relatively high-frequency optical modes, but its lack of a bandgap, presence of negative frequencies and structural instability limit its practical applications. In contrast, Low-Buckled Silicene (Fig. \ref{fig:1}(j)) shows a modified phonon dispersion due to its buckled geometry, which helps stabilize the structure while still exhibiting high-frequency phonon modes. Kagome Silicene (Fig. \ref{fig:1}(k)) introduces localized phonon modes due to the presence of flat electronic bands, which can influence thermal conductivity and electron-phonon interactions. In a kagome lattice, phonon modes are vibrational excitations of the atoms that can exhibit unique characteristics due to the lattice's geometric frustration, often featuring "flat" phonon bands with very little dispersion, meaning the phonon frequency remains nearly constant across a large range of momentum space, and are particularly susceptible to softening due to electron-phonon coupling, potentially leading to phenomena like charge density waves (CDWs); these flat phonon modes are often associated with the trigonal atoms within the kagome structure; hhk-Silicene (Fig. \ref{fig:1}(l)) exhibits a phonon dispersion that confirms its dynamic stability, as there are no negative-frequency modes, which would indicate structural instability. The combination of $\text{sp}^{2}$ and $\text{sp}^{3}$ hybridization in hhk-silicene leads to distinct phonon branches, influencing thermal transport properties. The relatively smooth dispersion in the acoustic and optical branches suggests efficient heat conduction and tunable phonon transport.

\subsubsection{Group velocities}

\begin{figure*}[t!]

        \begin{center}
            \includegraphics[width=15cm]{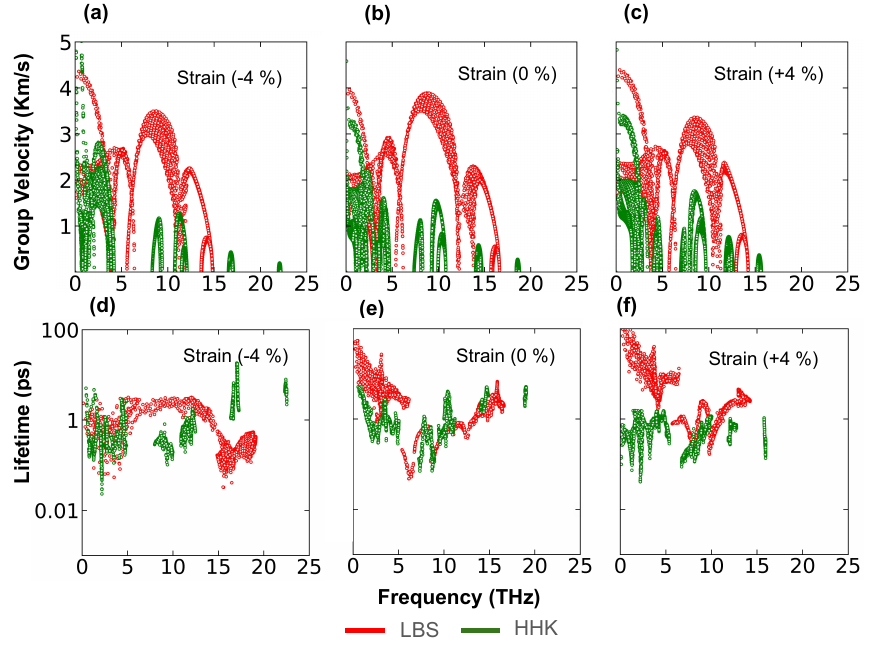}
            \caption{
            (a) - (c) : phonon group velocities for strains -4, 0 and 4 \% respectively for HHK Silicene. (d) - (f) : phonon lifetimes calculated using third-order force-constants for HHK Silicene.}
            \label{fig:3}
        \end{center}
        
\end{figure*}

The phonon group velocity ($V_g$) curves for Low-Buckled Silicene (LBS) and hhk-Silicene—are shown in Figure~\ref{fig:3} under three strain conditions: -4\%, 0\%, and 4\%. The phonon group velocity is a key parameter in thermal transport as it determines how efficiently heat is conducted through lattice vibrations.
\\

For all strain conditions, hhk-Silicene exhibits relatively high group velocities compared to other silicene topologies, indicating strong phonon transport and potential for effective heat dissipation in nanoelectronic applications. Under compressive strain (-4\%), shown in Figure~\ref{fig:3}(a), the phonon group velocity of hhk-silicene is slightly lower, suggesting that compressive strain reduces phonon propagation efficiency, which may lead to lower thermal conductivity. In the unstrained state (0\%), shown in Figure~\ref{fig:3}(b), hhk-silicene maintains a balanced phonon transport behavior, with velocities comparable to LBS but higher than Kagome and FS structures. Under tensile strain (4\%), shown in Figure~\ref{fig:3}(c), the group velocity increases, indicating enhanced phonon transport, which may improve thermal conductivity and heat dissipation capabilities.

\subsubsection{Relaxation times}
The phonon relaxation times for Low-Buckled (LBS) and hhk-Silicene—are presented in Figure~\ref{fig:3} as a function of phonon frequency under three strain conditions: -4\%, 0\%, and 4\%. Phonon relaxation time, which quantifies how long a phonon persists before scattering, plays a critical role in determining thermal conductivity, as longer lifetimes generally contribute to more efficient heat transport.
For all strain conditions, hhk-Silicene exhibits significantly longer phonon lifetimes compared to other silicene topologies, particularly in the low-frequency region ( 10 THz), indicating reduced phonon scattering and higher potential for thermal transport. Under compressive strain (-4\%), shown in Figure~\ref{fig:3} (d), phonon lifetimes for all silicene structures decrease slightly, suggesting increased phonon-phonon scattering, which may lead to lower thermal conductivity. In the unstrained state (0\%), shown in Figure~\ref{fig:3} (e), hhk-silicene maintains the longest phonon lifetimes, reinforcing its stable and efficient thermal transport properties. Under tensile strain (4\%), shown in Figure~\ref{fig:3} (f), phonon lifetimes increase slightly for hhk-silicene, indicating that tensile strain reduces scattering events, further enhancing thermal conductivity.

\begin{figure*}[t!]

        \begin{center}
            \includegraphics[width=15cm]{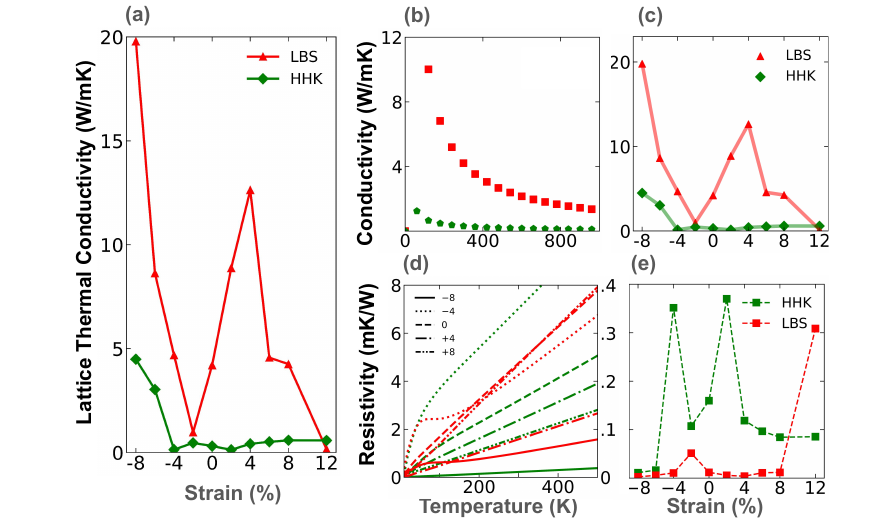}
            \caption{
            Transport properties of LBS and HHK silicenes. (a) : thermal conductivity plotted against strain (at 300 K). (b) : thermal conductivity plotted against temperature (at 0\% strain). (d) : thermal resistivity as a function of temperature. (e) : thermal resistivity as a function of strain.
            }
            \label{fig:4}
        \end{center}
        
\end{figure*}

\subsubsection{Lattice Thermal Conductivities}
The lattice thermal conductivity (LTC) of Low-Buckled (LBS) and hhk-Silicene is presented in Figure~\ref{fig:4}. The first plot (Figure~\ref{fig:4}a) shows the variation of LTC with strain at 300 K temperature, while the second plot (Fig. ~\ref{fig:4}b) illustrates LTC as a function of temperature at 0\% strain.
\\
The thermal conductivity of all silicene structures decreases with increasing temperature ($\text{T}$) due to enhanced phonon-phonon scattering. Among all structures, lbs-Silicene exhibits the highest LTC across all temperatures, suggesting superior heat transport efficiency. The hhk-topology structure shows the lowest LTC, indicating a more significant phonon scattering effect due to its unique lattice arrangement. The temperature-dependent trend follows the expected $1/\text{T}$ behavior, where thermal conductivity decreases as phonon scattering intensifies at higher temperatures.
\\
In Fig. ~\ref{fig:4}(a), LTC is plotted against strain at 300 K, showing that strain significantly influences thermal transport properties. Under compressive strain, LTC increases, likely due to an increase in bond strength and structural distortions which may lead to faster transfer of vibrations to release energy. Conversely, under tensile strain, LTC decreases, as a result of decaying spring constants between the atoms. 
\\
In Fig ~\ref{fig:4}(a), the LTC of hhk-silicene has great switching performance with respect to strain. The lattice thermal conductivity starts off at around 5 W/mK at -8\% strain and goes to near zero at -4\% and remains at that level. This trend can be exploited for thermal logic in strain-tronic applications.

\section{Discussion}

These results confirm that hhk-silicene exhibits good strain-dependent vibrational performance, making it an ideal candidate for strain-engineered thermal management applications. The ability to control lattice thermal conductivity through strain offers a promising pathway for integration in nanoelectronic and thermoelectric devices.


        

\section{Conclusion}
In this work, we have reported a comprehensive study of the vibrational properties of isotropically Silicene films of various topologies. The results obtained by first principles Density functional theory and Density functional perturbation theory point to the superior electronic properties of hhk-silicene which are at par with bulk-Si with a band gap of around 1.1 eV and a near linear tunability with strain. Also, hhk-silicene acts as a perfect strain-dependent thermal inverter as the shown in the figure below.

\section{acknowledgments}
We express our sincere gratitude to the IDAPT Computing Facility at the Indian Institute of Technology (IIT) BHU for providing their high-performance computing resources essential for this research.

\nocite{*}

\bibliography{cites}
\end{document}